\documentclass[fleqn,10pt]{wlscirep}
\usepackage[utf8]{inputenc}
\usepackage[T1]{fontenc}
\usepackage{soul}
\usepackage{siunitx}
\title{Asynchronous Event-Based Spectroscopy for Microsecond-Resolved Spectral Reconstruction}

\author[1, 2, *]{Joana M. Teixeira}
\author[1,2]{Tomás Lopes}
\author[1]{Tiago D. Ferreira}
\author[1]{Catarina S. Monteiro}
\author[1, 2]{Pedro A. S. Jorge}
\author[1, 2]{Nuno A. Silva}

\affil[*]{joana.m.teixeira@inesctec.pt}

\affil[1]{Center for Applied Photonics, INESC TEC, Rua do Campo Alegre 687, Porto, 4169-007, Portugal}

\affil[2]{Departamento de Física e Astronomia, Faculdade de Ciências da Universidade do Porto, Rua do Campo Alegre 687, Porto, 4169-007, Portugal}

\begin{abstract}
Many physical and chemical processes of interest evolve on timescales that push the limits of conventional spectroscopic instrumentation. Indeed, the temporal resolution of standard spectrometers is often insufficient to track these dynamics, which is connected to the fact that most systems rely on frame-based sensors, imposing fundamental constraints on acquisition speed, sensitivity, and data efficiency, frequently limiting practical operation to the kilohertz regime. In this work, we present an approach to circumvent this limitation by developing an event-based spectrometer to enable spectral reconstruction with microsecond temporal resolution by leveraging a Czerny–Turner configuration combined with asynchronous and event-driven sensing. A dedicated signal processing pipeline converts the resulting stream of binary events into calibrated spectra through temporal accumulation, geometric correction, and vertical spatial integration of the spectral line, covering a 234\,nm bandwidth in the visible range with a spectral resolution of approximately 0.18\,nm per pixel. Performance characterization under temporally modulated illumination demonstrates that the event-based spectrometer can reconstruct spectra at probing rates of up to tens of kilohertz, far exceeding the practical limits of a conventional frame-based spectrometer operated in parallel, while accurately preserving spectral peak positions and relative spectral features.
Finally, to further illustrate its potential applications, the system is validated in a microfluidic experiment integrated into an inverted microscope, where spectral changes induced by an absorbing dye are tracked with higher temporal fidelity and resolution comparing with the frame-based approach. These results establish event-based spectroscopy as a promising paradigm for real-time, high-temporal-resolution spectral measurements in dynamic and low-light applications.

\end{abstract}
\begin{document}

\flushbottom
\maketitle

\thispagestyle{empty}

\section*{Introduction}
Spectroscopy plays a central role in the study of light–matter interactions, providing access to material properties at molecular and atomic scales. Beyond static characterization, many physical, chemical, and biological processes are governed by time-dependent spectral variations, whose dynamics reveal key information about reaction pathways, energy transfer, and transient states \cite{Perez-Castillo2024, Hoghooghi2024}. Resolving these spectral changes in time is therefore essential for the investigation of rapidly evolving or short-lived phenomena. However, when addressing time-varying signals, conventional middle-range spectrometers based on charge-coupled devices (CCDs) or complementary metal–oxide–semiconductor (CMOS) sensors are strongly limited in acquisition speed, sensitivity, and dynamic range. These systems are typically constrained to the kilohertz regime, with temporal resolutions on the order of a few microseconds. For faster dynamics, specialized detectors such as intensified CCDs \cite{andor_iccd}, electron-multiplying CCDs \cite{evolve_emccd,pimax4}, or streak cameras \cite{wang2020single,takahashi1994new} can be employed, reaching temporal resolutions down to the picosecond range but at higher cost, reduced dynamic range, and increased system complexity. Alternatively, single-pixel photodetectors combined with wavelength-selective filters may be used to achieve high temporal resolution, at the expense of spectral bandwidth and the ability to monitor the full spectral evolution simultaneously \cite{Ghezzi2024}.

In this context, event-based sensors (EVS) represent a promising alternative for time-resolved spectroscopy, although their potential in this domain remains largely underexplored. Unlike traditional frame-based sensors, event-based cameras operate asynchronously, with each pixel independently reporting changes in logarithmic intensity that exceed a predefined threshold \cite{lenero2018applications}. This operating principle enables high temporal resolution and throughput, typically on the order of microseconds and equivalent megahertz event rates, while avoiding the redundancy inherent to frame-by-frame acquisition \cite{ cabriel2023event}. Such characteristics make event-based sensors particularly well-suited for capturing fast and transient optical phenomena.

The capabilities of event-based sensors have been extensively demonstrated in the context of high-speed vision tasks, including object tracking \cite{zhang2023frame, zhang2022spiking}, optical flow estimation \cite{benosman2013event, milde2015bioinspired}, and autonomous driving \cite{wan2021event}. However, only recently has their application begun to extend into photonics and optical measurement systems. Here, event-based cameras have been successfully used for high-speed particle tracking and velocimetry \cite{ni2012asynchronous, wang2020stereo,willert2022event}, where they replace conventional high-speed cameras while overcoming bandwidth and memory constraints. In advanced imaging systems, they have also enabled kilohertz-rate three-dimensional microscopy and ultrafast fluorescence measurements \cite{guo2023eventlfm}. Finally, and of particular relevance to the spectroscopy field, event-based cameras are also finding their applications, including differential spectral absorbance measurements via temporally modulated illumination\cite{takatani2021event}, and hyperspectral imaging via grating-based \cite{sopek2025development} or scanning wavelength solutions \cite{chen2026self}.

Yet, in spite of these advances, the use of event-based sensors for broadband spectroscopy has yet to be demonstrated. For such an application, the event-driven nature of these sensors is particularly relevant as it inherently suppresses constant illumination, generating events only in response to temporal changes in intensity. This property is especially advantageous in transmission measurements, where small spectral variations must often be detected against a strong and slowly varying background. Leveraging on all these advantages, this manuscript presents a novel spectrometer design that utilizes an event-based sensor to address the limitations of conventional frame-based systems. The spectrometer employs a Czerny–Turner configuration, combining standard dispersive optics with a high-resolution event-based vision sensor, and is integrated with a dedicated signal processing pipeline to convert asynchronous event streams into spectral profiles with microsecond-scale temporal resolution. The proposed system operates in the visible range (414.47 - 648.21\,nm) and is particularly suited for transmission measurements, where dynamic spectral changes can be monitored with high temporal fidelity. Moreover, the performance of the event-based spectrometer is characterized under temporally modulated illumination and benchmarked against a conventional frame-based spectrometer. Lastly, its applicability to dynamic measurements is demonstrated using a microfluidic experimental setup, where spectral changes induced by an absorbing medium are tracked in real time. These results illustrate the potential of event-based spectroscopy as a cost-effective and flexible approach for time-resolved spectral measurements, bridging the gap between conventional spectrometers and high-end ultrafast optical instrumentation.

\section*{Methodology}
\subsection*{Experimental Setup and spectrometer design}
\label{subsec:setup}

\begin{figure}[h!]
    \centering
    \includegraphics[width=\linewidth]{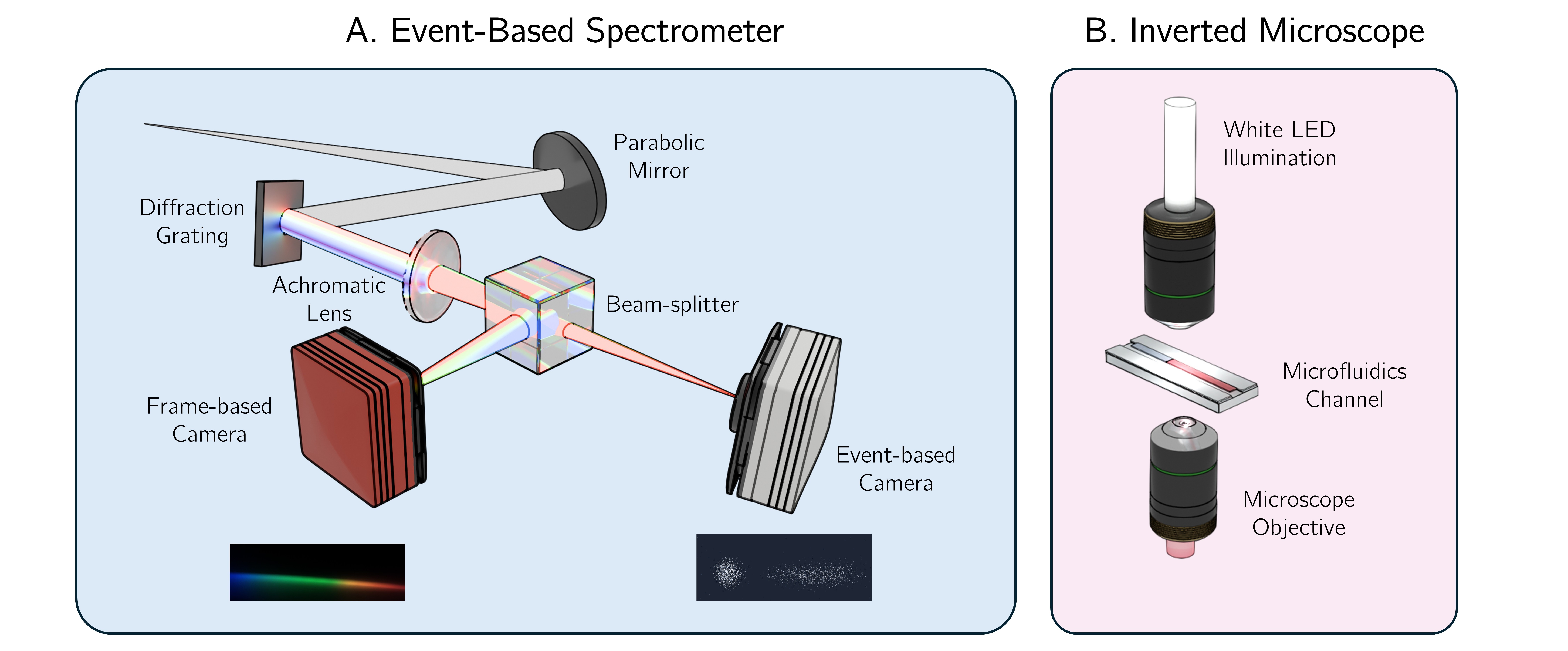}
    \caption{A. Illustration of the event-based spectrometer, consisting of a parabolic mirror for beam collimation, an achromatic lens for focusing, a beam splitter, and two detectors, namely an event-based camera used for spectral interrogation and a frame-based camera employed for monitoring and alignment. B. Illustration of the setup used for sample illumination and manipulation. The light transmitted through the sample and collected by the microscope objective is then sent to the event-based spectrometer system. }
    \label{fig:espectrometro}
\end{figure}

The experimental setup consists of two integrated components: a custom event-based spectrometer and a modified inverted microscope. The spectrometer uses a Czerny–Turner configuration with light entering the system through an optical fiber coupled to a \SI{50}{\micro\meter} entrance slit. An off-axis parabolic mirror with a focal length of 25.4\,mm (MPD119-P01, Thorlabs) collimates the beam after the slit, which is then dispersed by a reflective diffraction grating with a 500\,nm blaze wavelength (GR25-0605, Thorlabs) with a groove density of 600 grooves/mm. Finally, an achromatic lens with a 40\,mm focal length focuses the diffracted spectrum, through a beam-splitter, onto the sensing planes of an event-based vision sensor (EVK4 HD, Prophesee,  1280x720 pixels, 1/2.5'' sensor size) and a conventional camera (DCC1249C-HQ, Thorlabs, 1440x1080 pixels, 1/2.9'' sensor size ). This configuration is schematically represented in Figure \ref{fig:espectrometro}.

Coupled to the spectrometer, a modified inverted microscope provides controlled sample illumination and manipulation \ref{fig:espectrometro}B. A white LED illuminates the sample from above, with its position adjusted manually. The LED is modulated using a DAQ (USB-6351, National Instruments), which enables precise control over the waveform and frequency. The light is focused onto the sample using a 20× microscope objective, and the sample is positioned on translation stages that allow for fine spatial alignment. Transmitted light is then collected using a second 20× objective lens and focused using a 10\,cm focal length achromatic lens. A movable mirror directs the incoming light to two optical paths. The first path leads to a standard camera used for real-time sample visualization and position adjustments. The second path couples the transmitted light into a fiber optic that further splits the beam into two outputs: one routed to the event-based spectrometer and the other to a conventional spectrometer (AvaSpec‑ULS2048CL‑EVO, Avantes). The latter supports single-acquisition integration times down to \SI{30}{\micro\second} and continuous operation up to 1\,kHz. This configuration enables synchronized acquisition of spectral data from both spectrometers, allowing direct comparison between conventional and event-based optical sensing.

\subsection*{Signal Processing}
Event-based cameras operate on a per-pixel, asynchronous change-detection principle. Each pixel continuously tracks the incident radiance in the log-intensity domain and maintains an internal reference value for comparison. In theory, an event $e_k = (\textbf{x},t_k,p_k$) is triggered at pixel $\textbf{x} = (x,y)$ at time $t_k$ every time the intensity $I$ suffers a change that verifies
\begin{equation}
    |\log\left( I(x,y,t)\right) - \log\left( I(x,y,t-\Delta t_k)\right) | \geq C
\end{equation}
where $I(x,y,t-\Delta t_k)$ corresponds to the intensity at the last event detected at a given pixel, and $C$ to a threshold that can be controlled by the user.
The transmitted event packet then consists of the $x, y$ pixel location, a timestamp $t$ (typically with microsecond resolution), and a 1-bit polarity $p$ that encodes the sign of the change (brightness rise, +1 or reduction, -1). Consequently, the sensor output is inherently data-driven, with scenes with stronger or faster intensity variations producing higher event rates, whereas static regions generate few or no events.

To process the raw data acquired from the event-based camera, we define a signal processing pipeline that converts the stream of events into a usable spectrum. Instead of using the full pixel array of the camera, we restrict the acquisition to a region of interest (ROI), consisting of the full horizontal range (1280 pixels) and  150 pixels in the vertical direction. This reduces the amount of data to be stored, the computational overhead, and the expected pixel latency, which may bound overall performance\cite{sengupta2024demystifying}. The incoming stream of events is then accumulated, pixel-by-pixel, into temporal frames, each with a duration of a few tens of microseconds (typically between 20 and \SI{100}{\micro\second}).

\begin{figure}[h!]
    \centering
    \includegraphics[width=\linewidth]{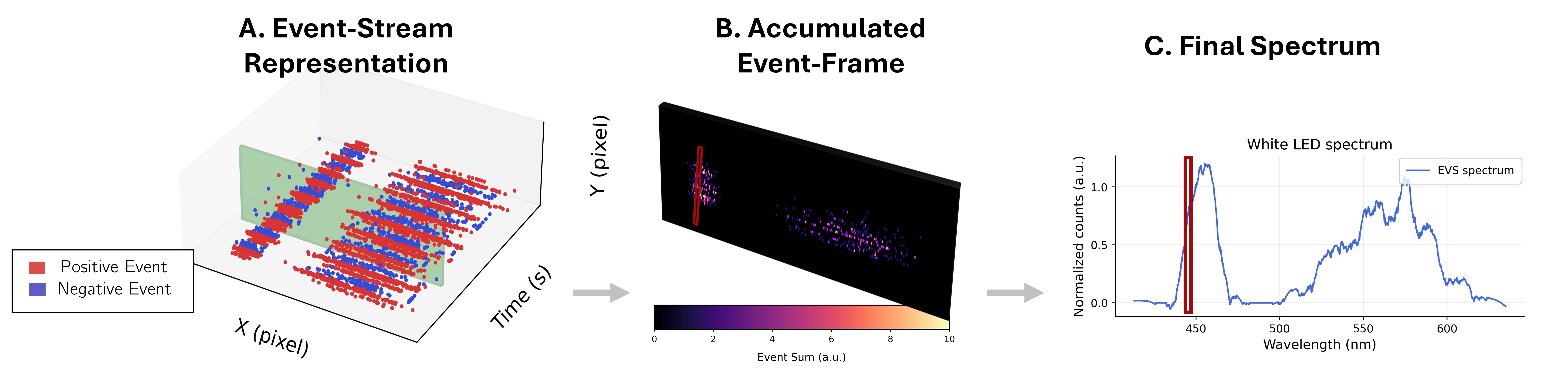}
    \caption{(A) Asynchronous event stream represented in the spatio-temporal domain, where each event encodes changes in intensity (red and blue corresponding to brightness increases and decreases, respectively). (B) Events accumulated over a short temporal window to form an event-frame. Brighter pixels correspond to a higher number of accumulated events. (C) The final spectrum is obtained by summing events column-by-column across the selected vertical region, yielding a one-dimensional spectral profile.}
    \label{fig:processamento}
\end{figure}

Because small misalignments in the optical setup can cause the dispersed beam to reach the camera at a slight angle, the frame undergoes first a computational tilt correction. For this, we fit a linear function to the distribution of pixel intensities, extract the resulting angle, and apply a rotation to realign the frames. From the corrected frames, we identify the horizontal line with the highest concentration of events, corresponding to the center of the focal region of the dispersed spectrum. Around this line, we select a band of $\pm 50$ rows and sum column by column across this region, producing a one-dimensional spectrum with 1280 points (see Figure \ref{fig:processamento}). Using multiple vertical detector rows increases the signal-to-noise ratio by coherently summing multiple independent measurements of the same spectral signal, a principle that is increasingly adopted in modern spectrometer architectures \cite{Dieter2022}. The event-based spectrometer naturally exploits this approach, unlike traditional single-line CCD spectrometers, by asynchronously integrating information across many vertical pixels with microsecond-scale latency.

Finally, to suppress high-frequency noise while preserving the underlying spectral structure, each spectrum is smoothed using a Savitzky–Golay filter with a window length of 50 points and a third-order polynomial. This smoothing step suppresses noise artifacts without distorting the underlying spectral profile. This step is especially important when using low-intensity illumination or very low temporal acquisition windows. 

In parallel, and for benchmarking purposes, the spectra acquired with the conventional spectrometer were also processed using a dedicated signal-conditioning pipeline applied to each recorded scan. First, the wavelength range was restricted to the region of interest by discarding low-signal edge pixels, reducing the original spectra to the spectral window relevant for the experiments. The resulting spectra were then smoothed to suppress noise and highlight key information. Finally, baseline distortions arising from background illumination and instrumental offsets can be removed using an asymmetric least-squares baseline correction algorithm. This step compensates for slowly varying or constant components of the signal, which are inherently suppressed by the event-based sensor, ensuring a consistent comparison between the two modalities. The resulting spectra are therefore well-suited for both qualitative and quantitative evaluation against the event-based reconstructions.

\subsection*{Spectrometer Calibration}
The event-based spectrometer can be calibrated in wavelength by using three reference light sources corresponding to red, green, and blue LEDs. For this process, each LED is driven in a pulsed regime and recorded sequentially by the event-based sensor, generating an asynchronous stream of events. The events are then reconstructed into temporal frames using a fixed temporal accumulation window, as previously described in the processing pipeline. In this case, a longer temporal integration window of 1\,ms should be selected, as the source signal remains consistent across multiple pulses, and a higher signal-to-noise ratio should be prioritized. For each LED, the spatial intensity distribution along the dispersion axis is then integrated to yield an event-based spectral profile.

In parallel, the same LEDs are also characterized using the conventional spectrometer under identical illumination conditions, providing reference spectra in absolute wavelength units. Both datasets are then normalized and compared to identify the spectral response corresponding to each color channel. A simplified linear wavelength–pixel mapping
\begin{equation}
\lambda_{fit} = a \times pixel + b
\end{equation}
can be fitted by minimizing the squared difference between the event-based and reference spectra through a least-squares optimization routine.
Note that in this model, the coefficient $a$ represents the spectral dispersion of the system (in nanometers per pixel), defining how wavelength varies across the sensor’s horizontal axis, while $b$ corresponds to the wavelength associated with the first active pixel in the reconstructed spectrum. To enable faster convergence, the optimization can be initialized with a manual approximation, estimated from the expected diffraction geometry and the relative position of the LED peaks observed in both the event-based and conventional spectra. This provides a physically reasonable starting point for the fitting routine, which then refines the parameters \textit{a} and \textit{b} to minimize the spectral misalignment between the two measurements.

\section*{Results and Discussion}

\subsection*{Spectrometer Characterization}

The spectrometer was first calibrated following the procedure described in the previous section in order to establish a connection between the event camera pixels and the corresponding wavelengths of the incoming light. From this calibration, the spectrometer was found to operate in the visible range, covering wavelengths from 414.47 to 648.21\,nm, which corresponds to a total spectral span of 233.7\,nm with a per-pixel spectral resolution of 0.18 nm. The calibrated central wavelength is approximately 530\,nm, which is consistent with the nominal design wavelength of the diffraction grating used in the system (500\,nm), confirming proper alignment of the dispersive optics. It is important to note that this spectral window is not intrinsic to the sensor itself but depends on the optical alignment of the diffracted light relatively to the detector. By adjusting this alignment, the operational range of the spectrometer can be shifted toward shorter or longer wavelengths, allowing different regions of the visible spectrum to be targeted depending on the application. The configuration reported in this manuscript, therefore, corresponds to an alignment chosen to capture a broad portion of the visible spectrum.

With the spectral operating range and resolution established through calibration, the spectrometer was subsequently evaluated under temporally modulated illumination to assess its response to rapidly varying optical signals. These experiments were designed to probe the limits of spectral reconstruction under high-frequency excitation while remaining within the calibrated spectral window and optical configuration described above. For this, a white LED was driven with a square-wave waveform generated by a DAQ system, enabling precise control of the modulation frequency and duty cycle. The duty cycle was set to 0.7 to maximize the delivered optical power, and the drive amplitude was fixed at the DAQ limit of 10\,V. It should be noted that the LED driver did not provide independent current control and had limited modulation bandwidth. As the modulation frequency increased, the finite rise and fall times of the drive signal prevented the LED current from reaching its steady-state level within each pulse. This reduced the time-averaged drive current and, consequently, the effective optical power reaching the detector. As a consequence, this effect is reflected in the measured signal levels at higher modulation frequencies.

Two modulation frequencies were selected to perform a detailed evaluation: 30\,kHz and 40\,kHz. Owing to the polarity-sensitive nature of the event-based sensor, positive and negative events can be used independently to detect the rising and falling edges of the square-wave modulation. As a consequence, the previous modulation frequencies correspond to effective spectral probing rates of 60\,kHz and 80\,kHz, respectively. This demonstrates the capability of the event-based spectrometer to interrogate the optical signal at temporal scales beyond those accessible to conventional middle-range spectrometers, which typically lie within the 1\,kHz range with limited buffer memory. For comparison, our frame-based spectrometer was operated in parallel. The integration time was set to 500\,µs, which in principle should allow probing frequencies up to approximately 1\,kHz, following Nyquist's theorem. However, in practice, limitations associated with data transfer and storage resulted in an effective acquisition rate of approximately 500\,Hz for the model under consideration. Under these conditions, it was not possible for the frame-based spectrometer to detect a single cycle of the LED modulation at either of the test frequencies.

\begin{figure}[h!]
    \centering
    \includegraphics[width=\linewidth]{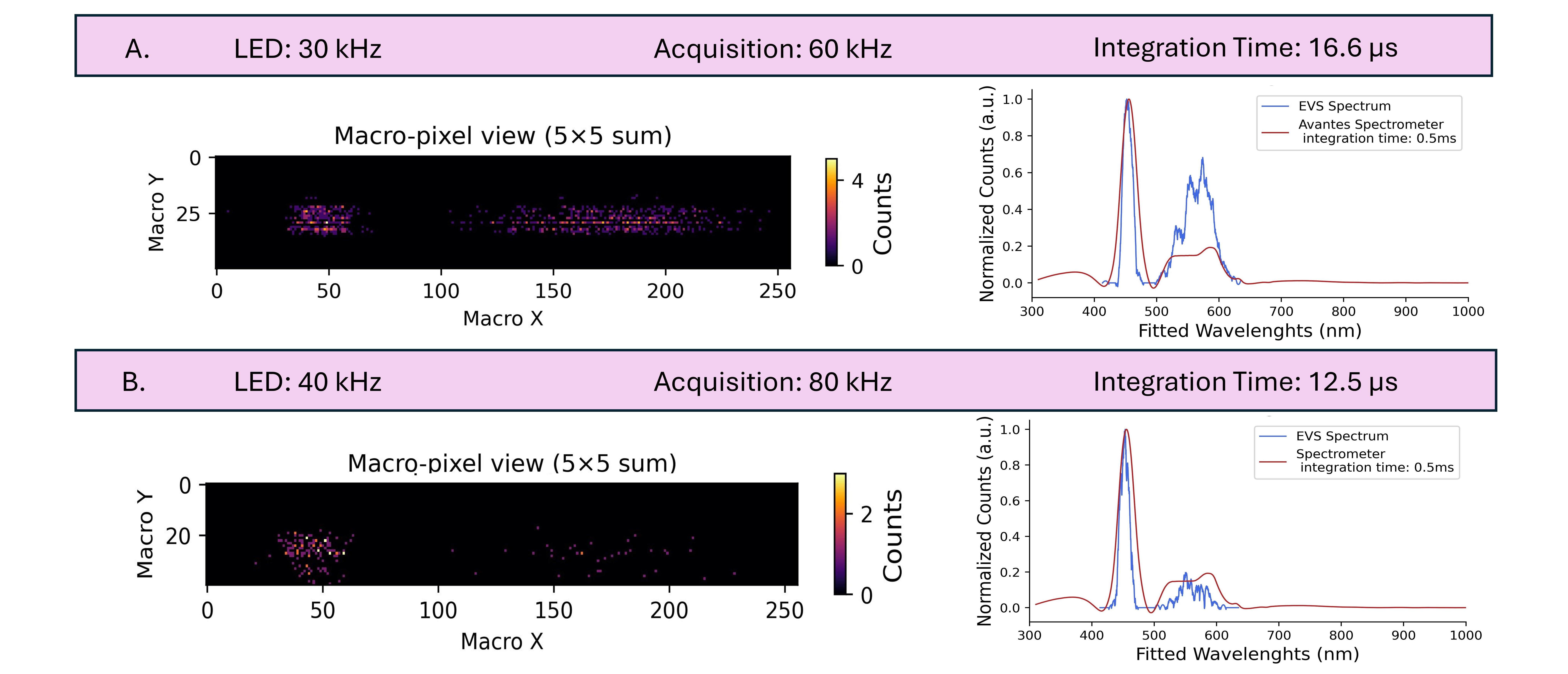}
    \caption{Accumulated event image and resulting spectra for a white LED modulated at 30 and 40\,kHz (cases A. and B.). The event image was constructed with macropixels to improve visibility. Note that spatial binning is applied only for visualization, and all spectra are reconstructed from the original, full-resolution event data. For both spectra, a reference line is presented, in red, obtained from a commercial spectrometer.}
    \label{fig:speed_caractherization}
\end{figure}

\begin{figure}[h!]
    \centering
    \includegraphics[width=0.85\linewidth]{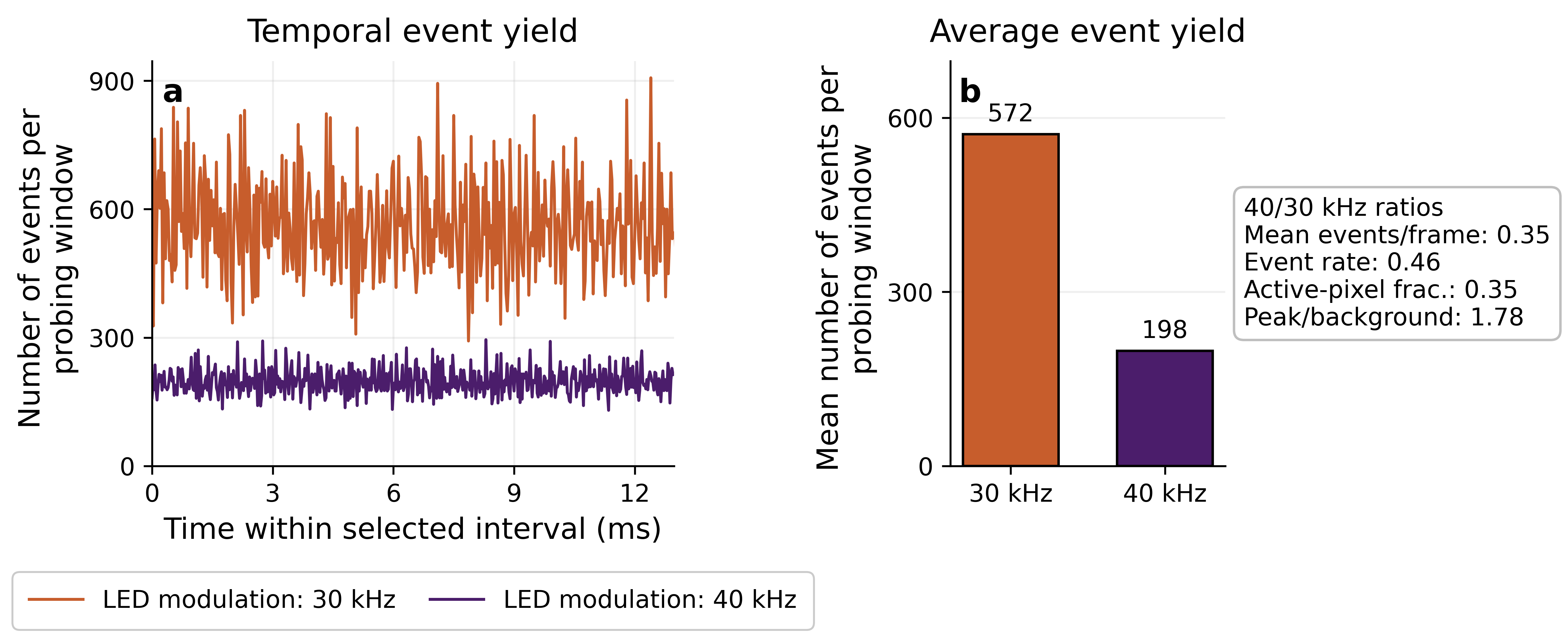}
    \caption{Event-yield comparison for white-LED modulation at 30 and 40 kHz. (a) Number of detected events per probing window over a representative temporal interval, where each point corresponds to the total number of events accumulated within one modulation period (integration time). (b) Mean number of events per probing window for each modulation frequency. The inset box reports the ratios between the 40 kHz and 30 kHz conditions for several metrics: mean events per frame, defined as the average number of events detected within one probing window; event rate, obtained by normalizing the event count by the integration time (events per second); active-pixel fraction, defined as the proportion of pixels that generated at least one event in the accumulated frame; and peak-to-background ratio, computed as the ratio between the event density within the signal band and that of the remaining background region. The reduction in event yield at 40 kHz is consistent with the lower optical power reaching the sensor at higher modulation frequencies.}
    \label{fig:event_number}
\end{figure}

The experimental results for this approach are summarized in Figure \ref{fig:speed_caractherization}. For each modulation frequency, the figure reports the LED modulation frequency, the effective probing rate, and the integration time used by each sensor. Notably, due to the asynchronous nature of the event-based camera, both rising and falling intensity transitions generate events, resulting in an effective probing rate that is twice the LED modulation frequency. These quantitative indicators are followed by visualizations of the accumulated event frames, constructed by summing and spatially binning the original event data into 5×5 macro-pixels, to simplify their visualization. As expected, the accumulated event frame corresponding to 30\,kHz modulation exhibits a higher event density than that obtained at 40\,kHz, consistent with the reduced optical power reaching the sensor at higher modulation frequencies. The reconstructed spectra obtained from the event-based spectrometer are shown alongside the corresponding spectra acquired with the commercial spectrometer. In both modulation cases, the event-based spectrometer accurately reproduces the positions of the dominant peaks and the relative spectral regions of interest. While absolute intensity information is not preserved in the normalized representation, the agreement in spectral shape confirms the fidelity of the event-based reconstruction at high temporal rates. 

The event yield as a function of time and modulation frequency is also summarized in Fig.\ref{fig:event_number}. Figure \ref{fig:event_number}(a) shows the temporal evolution of the number of detected events per probing window, where each point corresponds to the total number of events accumulated within one modulation period. The signal remains stable over time, with a consistently higher event count observed at 30\,kHz compared to 40\,kHz. This difference is further quantified in Fig.~\ref{fig:event_number}(b), which reports the mean number of events per probing window. The 30\,kHz condition yields 
572$\pm$121 events per frame, whereas the 40 kHz condition produces 198$\pm$29, corresponding to a reduction of approximately  65\%. The figure further summarizes the relative changes in event rate, active-pixel fraction, and peak-to-background ratio, confirming a systematic decrease in detected signal at higher modulation frequencies. This behavior is consistent with the reduced optical power reaching the sensor as the modulation frequency increases.

Finally, it is worth noting that if the event-based camera were operated as a single-line detector, its asynchronous binary response would lead to equal peak heights for spectrally distinct features, as each wavelength would generate similar event activity once threshold crossings occur. By exploiting multiple vertical pixels per wavelength, the proposed system overcomes this limitation: increased optical power results in a broader vertical spread of events, yielding a higher accumulated event count. This effect enables meaningful comparison of relative peak strengths across the spectrum while preserving the high temporal resolution intrinsic to event-based sensing.

\subsection*{Tracking spectral changes on a Microfluidics System}
To demonstrate the applicability of the event-based spectrometer to monitor dynamic processes, its performance in tracking spectral changes within a microfluidic system was also investigated. The experiments were conducted using a microfluidic channel integrated into the inverted microscope described in the experimental setup section. Initially, the channel was filled with a transparent medium, so the measured transmission provided a baseline spectrum of the illumination source, with negligible sample-induced attenuation or spectral reshaping. During these experiments, a red food dye was introduced into the channel, inducing a change in the transmitted spectrum as the absorbing medium progressed through the field of view.

Spectral datasets were acquired simultaneously using the event-based and conventional frame-based spectrometers. For the event-based camera, a region of interest was, once more, defined to restrict acquisition to the spectrally dispersed region. A square-wave trigger signal (3.3\,V amplitude, 0.5 duty cycle) generated by the data acquisition system was supplied in parallel to both the event-based system and the frame-based spectrometer to ensure temporal synchronization. Note, however, that this trigger signal does not affect the event-stream acquisition, instead registering an external trigger event timestamped at each rising edge. 

For the frame-based spectrometer, we experimentally determined that the maximum trigger rate at which all triggers could be reliably registered was 250\,Hz. Accordingly, an acquisition time of 2\,ms was selected, which lies within this operational limit and provides sufficient exposure for spectral measurements. To maximize the optical power transmitted through the microfluidic channel, the illumination LED was modulated at 1\,kHz throughout the experiment.

To detect changes in the medium within the microfluidic channel, event-based frames were constructed at an effective rate of 2\,kHz by accumulating both positive and negative events. The total number of events per frame was then evaluated as a proxy for the transmitted optical power. As the red dye enters the channel, increased absorption leads to a reduction in the light reaching the sensor, which is reflected in a decrease in the total event count per frame. Owing to the high temporal resolution of the event-based acquisition, individual illumination cycles can be resolved, enabling precise temporal localization of the onset of spectral changes.

\begin{figure}[h!]
    \centering
    \includegraphics[width=\linewidth]{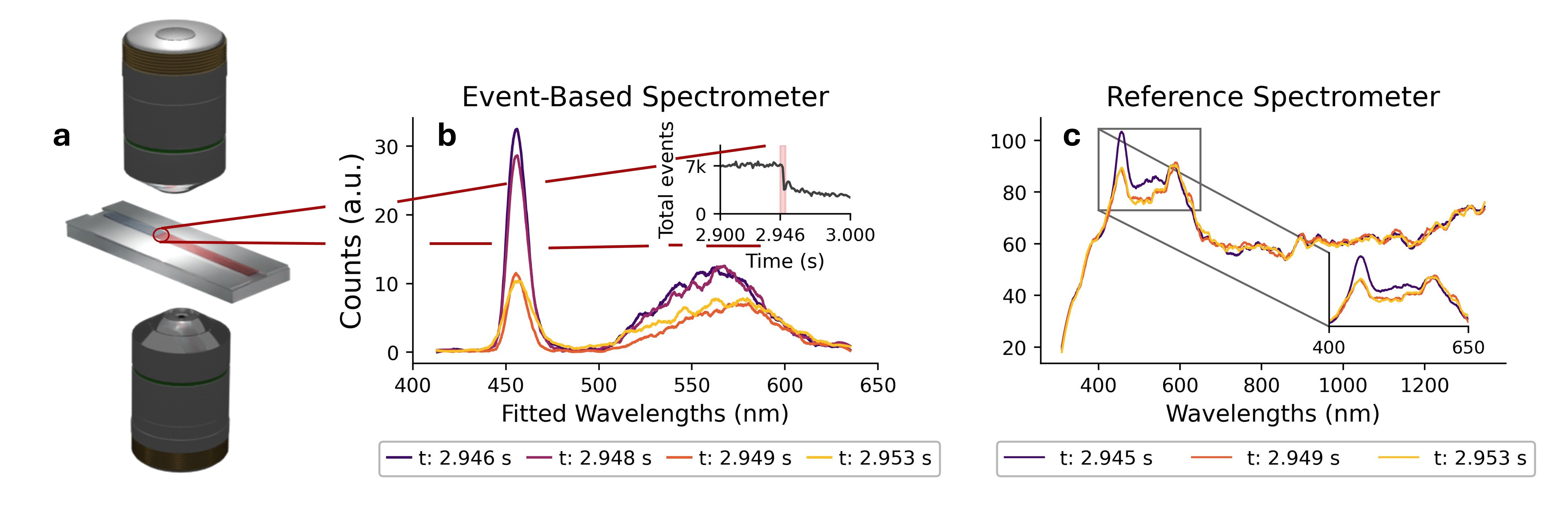}
    \caption{Spectral tracking during a microfluidic experiment in an inverted microscope configuration (a). (b) Spectra reconstructed from the event-based spectrometer (EVS) at selected time instants, with colors encoding acquisition time. The inset shows the total number of events per frame, highlighting the moment of spectral change induced by the introduction of the absorbing medium (highlighted in red). (c) Corresponding spectra acquired with a reference frame-based spectrometer at matching times. The inset highlights the EVS spectral window (400–650\,nm) within the full measurement range of the reference device. Colors are matched across both panels according to the closest acquisition timestamps.}
    \label{fig:tracking}
\end{figure}

The results of the microfluidic tracking experiment are presented in Figure \ref{fig:tracking}. The figure compares the spectra reconstructed from the event-based spectrometer (b) with those acquired using a conventional frame-based spectrometer (c) during the introduction of the dye into the microfluidic channel (a). In both cases, the spectral modification induced by the changing medium is clearly detected, confirming that both approaches are capable of tracking variations in the transmitted spectrum. However, the event-based spectrometer provides a substantially higher temporal sampling during the transition. As shown in Fig.\ref{fig:tracking}b, multiple intermediate spectra are reconstructed over the duration of the change, capturing the progressive attenuation of the blue spectral peak. This fine temporal resolution is not accessible with the frame-based spectrometer under the same conditions, where only a limited number of spectra are recorded during the transition.

Additionally, note that the inherent event-driven operation directly suppresses contributions from constant illumination, effectively enhancing spectral contrast under low-light conditions. This leads to improved identification of spectral features, particularly in regions of reduced signal, when compared to the frame-based acquisition. In this experiment, the spectra obtained from the reference spectrometer are intentionally presented without additional post-processing, enabling a direct comparison between both approaches. As demonstrated by the characterization experiments discussed in the previous section, operation under higher illumination levels would allow the event-based spectrometer to achieve analogous results at substantially higher effective frame rates.


Results for transmission of a cuvette configuration are also presented in the Supplementary Material (Section S1, Fig.S1), as well as additional experiments involving multiphase transport that are detailed in the Supplementary Material (Section S2, Figs. S2 and S3). In particular, for the latter, a transparent carrier fluid (index-matching oil) containing discrete inclusions of red dye was driven through the microfluidic channel, forming localized colored bubbles. As these dye-rich regions traversed the optical interrogation zone, they induced rapid, spatially confined spectral perturbations. The event-based spectrometer was able to capture these transient variations, demonstrating sensitivity to localized compositional heterogeneity within the flow. Although a precise temporal correlation between individual inclusions and reference spectral signatures was not possible in the current implementation due to the absence of synchronized acquisition, it is possible to observe the wavelength shifts corresponding to the detection of these bubbles, highlighting the potential of the approach for high-speed monitoring of complex microfluidic systems. All in all, and together with the results previously presented in this section, these demonstrate the feasibility and advantages of event-based spectroscopy for real-time monitoring of dynamic processes, establishing this experiment as a proof-of-concept for high-temporal-resolution spectral tracking in microfluidic systems.

\subsection*{Real-time Interrogation}

In addition to the previously described signal processing pipeline, the proposed system can leverage the intrinsic advantages of event-driven sensing to perform a computationally efficient real-time operation. Indeed, real-time spectral reconstruction can be achieved by accumulating events over short temporal windows and projecting them directly onto the spectral axis. Events are accumulated computationally along the vertical dimension of the sensor, corresponding to the spatially dispersed spectral line, resulting in a one-dimensional intensity profile as a function of pixel position. An example video showcasing the interface developed to monitor this real-time operation can be accessed \href{https://drive.google.com/file/d/1XCPo5wDXlBb4blV_YRtYp5TXJI5PPSsL/view?usp=drive_link}{here}. We shall note also that the asynchronous operation of each pixel may allow us to bypass (up to some level) the effective pixel-level latency of the sensor, which is on the order of $\sim$\SI{100}{\micro\second}, enabling spectral updates at timescales dictated by the chosen event accumulation window rather than by full-frame exposure and readout constraints.
In this demonstration, we can see both the generated event frames and spectra as a result of, first, ambient light variations and, later, the modulated white LED used in the previous tests.

A key advantage of the event-based formulation is that spectral estimation can be made conditional on the presence of meaningful activity. In practice, a minimum event-count threshold can be defined for each temporal window. If the number of detected events falls below this threshold, the system infers that no significant spectral changes have occurred. So, when the observed optical signal remains static, no events are produced, and no data is processed or stored. This conditional execution further reduces computational load and reinforces the sparse, data-driven nature of the approach, in contrast to frame-based systems that would continue to acquire and store identical spectra at every integration period. Besides, the region of interest of the event-based sensor can also be dynamically configured to match the spectral features of interest. Rather than monitoring the full spectral range, the ROI may be restricted to specific wavelength intervals corresponding to known emission bands or spectral lines. This spatial restriction directly reduces the number of generated events and, consequently, the data throughput and processing requirements. 


\section*{Conclusion}

In this work, we presented a novel event-based spectrometer architecture and demonstrated its applicability to high-speed and dynamic spectral measurements. By combining a Czerny–Turner optical configuration with an event-based vision sensor, we showed that asynchronous sensing can be effectively leveraged to reconstruct spectra with high temporal resolution while maintaining a compact and flexible optical design. A complete signal processing pipeline was developed to convert streams of events into calibrated spectral profiles, including spatial correction, vertical summation for signal-to-noise enhancement, and wavelength calibration against a reference spectrometer.

Through a detailed spectrometer characterization, we established the operational spectral range, resolution, and temporal capabilities of the system. Experiments with high-frequency LED modulation demonstrated that the event-based spectrometer can probe optical signals at effective rates of tens of kilohertz, far exceeding the practical limits of a conventional frame-based spectrometer operating under comparable conditions. Despite operating under reduced illumination at higher modulation frequencies, the event-based reconstructions accurately preserved spectral peak positions and relative spectral features, validating the fidelity of the approach.

The applicability of the system to dynamic measurements was further demonstrated through a microfluidic tracking experiment, where spectral changes induced by the introduction of an absorbing dye were monitored in real time. While both the event-based and frame-based spectrometers successfully detected the spectral transition, the event-based approach provided significantly denser temporal sampling during the change. This enabled the capture of intermediate spectral states, particularly revealing the progressive attenuation of the blue spectral component. Moreover, the intrinsic suppression of constant illumination by the event-driven sensor enhanced spectral contrast under low-light conditions, facilitating improved peak identification without additional post-processing. These results establish the proposed system as a proof-of-concept for high-temporal-resolution spectral tracking in microfluidic and other dynamic environments.

Overall, the real-time operation of the event-based spectrometer demonstrates how event-driven sensing enables low-latency, adaptive, and resource-efficient spectral measurements. By exploiting temporal sparsity, bypassing unnecessary pixel-level delays, conditionally executing spectral reconstruction, and restricting processing to relevant spatial regions, the system achieves responsive spectral interrogation while minimizing memory usage and computational overhead when compared to conventional frame-based spectroscopic systems. This work highlights the potential of event-based spectroscopy as a complementary paradigm to traditional spectrometers, particularly for applications requiring high temporal resolution, efficient data handling, and real-time responsiveness.

\section*{Acknowledgments}
Joana M. Teixeira and Tomás Lopes acknowledge the support of the Foundation for Science and Technology (FCT), Portugal, through Grants 2024.00426.BD and 2024.01830.BD, respectively. Nuno A. Silva acknowledges the support of FCT under the grant 2022.08078.CEECIND/CP1740/CT0002 \\(https://doi.org/10.54499/2022.08078.CEECIND/CP1740/CT0002). Tiago D. Ferreira acknowledges the support of FCT under the grant 2024.10684.CEECIND

\section*{Data availability}
The data and code used in the production of this manuscript can be made available upon reasonable request.

\bibliography{sample_2}

\end{document}